# Fiber-guided modes conversion using superposed helical gratings

Yancheng Ma[1], Liang Fang[2], Guoan Wu[1*]

[1]School of Optical and Electronic Information, Huazhong University of Science and Technology, Wuhan 430074, Hubei, P.R.China

[2]Wuhan National Laboratory for Optoelectronics, School of Optical and Electronic Information, Huazhong University of Science and Technology, Wuhan 430074, Hubei, P.R.China

*Corresponding author: GuoanWu_HUST@163.com

**Abstract:** Optical fibers can support various modal forms, including vector modes, linear polarization (LP) modes, and orbital angular momentum (OAM) modes, etc. The modal correlation among these modes is investigated via Jones matrix, associated with polarization and helical phase corresponding to spin angular momentum (SAM) and OAM of light, respectively. We can generate different modal forms by adopting superposed helical gratings (SHGs) with opposite helix orientations. Detailed analysis and discussion on mode conversion is given as for mode coupling in optical fibers with both low and high contrast index, respectively. Our study may deepen the understanding for various fiber-guided modes and mode conversion among them via fiber gratings.

## 1. Introduction

The light field has some inherent properties, such as its intensity, wavelength, polarization, and phase. The latter two have attracted a rapidly growing interest in recent years, due to their unique features. The polarization has become extremely diversiform with exploiting arbitrary states of polarization [1, 2], not just presents in the well-known vector fields with abundant vectorial polarization, such as cylindrically symmetric beams [3]. It can yield the optical chirality and spin, and manifests intrinsic spin angular momentum (SAM) of field [4]. As for the phase, its gradient of light beams brings about optical vortices producing orbital angular momentum (OAM) [4-6]. These features of light field or beam have been opening novel applications in various realms, such as electron acceleration [7], optical trapping [8, 9], laser machining [10], three-Dimensional focus engineering [11], mode-division multiplexing (MDM) [12, 13] and unidirectional excitation of surface and waveguide modes [14] in optics communications, and quantum optics and information [15], etc. Optical modes carrying OAM is characterized by the term of $\exp(il\phi)$ with $l$ being the topological charge number [5]. In optical fibers, modes carrying both SAM and OAM can be attributed to the combination of two vector modes [16] that are supplied with vector Helmholtz equation [17].

Optical modes in common weakly guiding fibers (WGFs) exhibit in the form of linear polarization (LP) modes due to the effect of mode degeneracy [18]. However, in the high-contrast-index fibers that have high refractive index difference between fiber core and cladding, LP modes would split into the vector mode components [17, 18], because of not equal effective index. The vector modes are the true eigenmodes of optical fibers, and the even and odd ones with a $\pi/2$ phase shift can combine into orbital angular momentum (OAM) modes. Conversely, two OAM modes with the same signs of SAM and OAM can combine into HE vector modes, whereas those with the opposite signs form EH modes [19]. Therefore, both vector and OAM modes can be respectively regarded as complete mode bases to combine fiber modes in different forms. The correlation of these model forms is investigated via the methods of Jones matrix in detail in the beginning of this article. Apart from three kinds of common fiber modes discussed above, we find that the WGFs also support other modal forms, such as OAM modes with linear polarization, and the modes with the states of circularly polarized lobes (CL) or hybrid polarization (HP).

Furthermore, based on the coupling principle of helical gratings (HGs) [20], we propose the superposed helical gratings (SHGs) to generate fiber-guided modes in different modal forms. This method of mode conversion is analogous to generation high order modes using tilted fiber gratings[21, 22]. Here we supervise the angular momentum states in the process of mode coupling between different modal forms through the SHGs. We believe that our investigation on mode correlation and conversion in this article provides a clear perspective on the understanding of the relationship among different modal forms, and it may make sense to exploit the angular momentum of these modes with application to optical trapping, mode manipulation, and optics communications, etc.

## 2. Mode decomposition via Jones matrix

In this section, we discuss the OAM mode components of fiber eigenmodes and manifest them via Jones matrix, and reveal the mode correlation between different fiber-guided modal forms. First of all, the electric field of eigenmodes need to be expressed in Cartesian coordinate system. The transformation relation between this coordinate system and the polar coordinate system is as follows:

$$\begin{bmatrix} e_x \\ e_y \end{bmatrix} = \begin{bmatrix} \cos\phi & -\sin\phi \\ \sin\phi & \cos\phi \end{bmatrix} \cdot \begin{bmatrix} e_r \\ e_\phi \end{bmatrix}. \quad (1)$$

The hybrid modes ($HE_{mn}/EH_{mn}$) can be transferred as

$$HE_{mn}/EH_{mn} = \sqrt{\frac{2}{1+\eta^2}} \cdot F_{mn}(r) \cdot \begin{bmatrix} \cos\phi & -\sin\phi \\ \sin\phi & \cos\phi \end{bmatrix} \cdot \begin{bmatrix} \cos m\phi \\ \eta \sin m\phi \end{bmatrix}, \quad (2)$$

and the radially and azimuthally polarized modes ($TM_{0n}$ and $TE_{0n}$) [23],

$$TM_{0n} = F_{0n}(r) \cdot \begin{bmatrix} \cos\phi & -\sin\phi \\ \sin\phi & \cos\phi \end{bmatrix} \cdot \begin{bmatrix} 1 \\ 0 \end{bmatrix}, \quad (3)$$

$$TE_{0n} = F_{0n}(r) \cdot \begin{bmatrix} \cos\phi & -\sin\phi \\ \sin\phi & \cos\phi \end{bmatrix} \cdot \begin{bmatrix} 0 \\ 1 \end{bmatrix}, \quad (4)$$

where $\eta$ indicates the proportion of azimuthal to radial field components for $HE_{mn}/EH_{mn}$ modes. It takes negative sign for $HE_{mn}$ modes, and positive sign for $EH_{mn}$ modes. $F_{mn}(r)$ and $F_{0n}(r)$ correspond to the radial-dependent distribution of electric fields of these modes. Using the decomposition relations,

$$\begin{bmatrix} \cos\phi & -\sin\phi \\ \sin\phi & \cos\phi \end{bmatrix} = \frac{1}{2} e^{i\phi} \begin{bmatrix} 1 & i \\ -i & 1 \end{bmatrix} + \frac{1}{2} e^{-i\phi} \begin{bmatrix} 1 & -i \\ i & 1 \end{bmatrix}, \quad (5)$$

$$\begin{bmatrix} \cos m\phi \\ \mp \sin m\phi \end{bmatrix} = \frac{1}{2} e^{im\phi} \begin{bmatrix} 1 \\ \pm i \end{bmatrix} + \frac{1}{2} e^{-im\phi} \begin{bmatrix} 1 \\ \mp i \end{bmatrix}. \tag{6}$$

The Jones vectors of fiber eigenmodes can be written as:

$$\mathrm{HE}_{mn} = \frac{1}{2} F_{mn}(r) \left\{ e^{i(m-1)\phi} \begin{bmatrix} 1 \\ i \end{bmatrix} + e^{-i(m-1)\phi} \begin{bmatrix} 1 \\ -i \end{bmatrix} \right\}, \tag{7}$$

$$\mathrm{EH}_{mn} = \frac{1}{2} F_{mn}(r) \left\{ e^{i(m+1)\phi} \begin{bmatrix} 1 \\ -i \end{bmatrix} + e^{-i(m+1)\phi} \begin{bmatrix} 1 \\ i \end{bmatrix} \right\}, \tag{8}$$

$$\mathrm{TM}_{0n} = \frac{1}{2} F_{0n}(r) \left\{ e^{i\phi} \begin{bmatrix} 1 \\ -i \end{bmatrix} + e^{-i\phi} \begin{bmatrix} 1 \\ i \end{bmatrix} \right\}, \tag{9}$$

$$\mathrm{TE}_{0n} = \frac{1}{2} i F_{0n}(r) \left\{ e^{i\phi} \begin{bmatrix} 1 \\ -i \end{bmatrix} - e^{-i\phi} \begin{bmatrix} 1 \\ i \end{bmatrix} \right\}. \tag{10}$$

Assuming that $\mathrm{HE}_{mn}/\mathrm{EH}_{mn}$ modes expressed above correspond to even modes, the corresponding odd modes with a $\pi/2$ angular offset can be given by

$$\mathrm{HE}_{mn}^{o} = \frac{1}{2} i F_{mn}(r) \left\{ e^{-i(m-1)\phi} \begin{bmatrix} 1 \\ -i \end{bmatrix} - e^{i(m-1)\phi} \begin{bmatrix} 1 \\ i \end{bmatrix} \right\}, \tag{11}$$

$$\mathrm{EH}_{mn}^{o} = \frac{1}{2} i F_{mn}(r) \left\{ e^{-i(m+1)\phi} \begin{bmatrix} 1 \\ i \end{bmatrix} - e^{i(m+1)\phi} \begin{bmatrix} 1 \\ -i \end{bmatrix} \right\}. \tag{12}$$

One can see that $\mathrm{HE}_{mn}$ modes comprises two OAM modes with the same sign of SAM and OAM that is spin-orbit aligned, whereas $\mathrm{EH}_{mn}$ modes comprises two OAM modes with the opposite sign of these angular momentum states that is spin-orbit dis-aligned [19]. It is obvious that OAM modes can be obtained by combination of even and odd vector modes with a $\pi/2$ phase shift, i. e.,

$$\mathrm{OAM}_{\pm m,n}^{\pm 1} = \mathrm{HE}_{m+1,n}^{e} \pm i \cdot \mathrm{HE}_{m+1,n}^{o} = F_{mn}(r) \cdot e^{\pm im\phi} \begin{bmatrix} 1 \\ \pm i \end{bmatrix}, \tag{13}$$

$$\mathrm{OAM}_{\pm m,n}^{\mp 1} = \mathrm{EH}_{m-1,n}^{e} \pm i \cdot \mathrm{EH}_{m-1,n}^{o} = F_{mn}(r) \cdot e^{\pm im\phi} \begin{bmatrix} 1 \\ \mp i \end{bmatrix}, \tag{14}$$

for $m \geq 2$, and

$$\mathrm{OAM}_{\mp 1,n}^{\pm 1} = \mathrm{TM}_{0n} \pm i \cdot \mathrm{TE}_{0n} = F_{mn}(r) \cdot e^{\mp i\phi} \begin{bmatrix} 1 \\ \pm i \end{bmatrix}, \tag{15}$$

for $m = 1$.

In WGFs, it is well known that the mode pairs of $\mathrm{HE}_{m+1,n}$ and $\mathrm{EH}_{m-1,n}$ ($m \geq 2$) modes and mode group of $\mathrm{HE}_{21}$, and $\mathrm{TM}_{01}$, $\mathrm{TE}_{01}$ modes would degenerate into LP modes, due to the approximate effective index. Therefore, it is obvious that the mode degeneration may also occur among the corresponding OAM modes. When the four kinds of OAM modes are set as a group of complete bases, arbitrary fiber modes (AFMs) can be generated as

$$\mathrm{AFM} = S_1 \cdot \mathrm{OAM}_{+m,n}^{+1} + S_2 \cdot \mathrm{OAM}_{-m,n}^{-1} + S_3 \cdot \mathrm{OAM}_{+m,n}^{-1} + S_4 \cdot \mathrm{OAM}_{-m,n}^{+1} \tag{16}$$

The usual mode combinations are listed in Tab. 1. As the expressions using Jones Matrix, the OAM modes with linear polarization in the $x$ direction are

$$\mathrm{OAM}_{\pm m,n}^{x} = 2 F_{mn}(r) \cdot e^{\pm im\phi} \begin{bmatrix} 1 \\ 0 \end{bmatrix}, \tag{17}$$

and even and odd LP modes in the $x$ direction correspond to

$$\mathrm{LP}_{mn}^{e,x} = F_{mn}(r)\cdot\cos m\phi \begin{bmatrix} 1 \\ 0 \end{bmatrix}, \tag{18}$$

$$\mathrm{LP}_{mn}^{o,x} = F_{mn}(r)\cdot\sin m\phi \begin{bmatrix} 1 \\ 0 \end{bmatrix}. \tag{19}$$

In addition, there are other modal forms characterized by circularly CL states

$$\mathrm{CL}_{mn}^{e,\pm 1} = 2F_{mn}(r)\cdot\cos m\phi \begin{bmatrix} 1 \\ \pm i \end{bmatrix}, \tag{20}$$

and HP states

$$\mathrm{HP}_{mn}^{e,\pm 1} = F_{mn}(r)\cdot\cos m\theta \begin{bmatrix} 1 \\ \pm i \end{bmatrix} + i\cdot F_{mn}(r)\cdot\sin m\theta \begin{bmatrix} 1 \\ \mp i \end{bmatrix}, \tag{21}$$

**Tab. 1 Mode combination with OAM mode bases**

| $(S_3, S_4)$ | $(S_1, S_2)$ | $(1,1)$ | $(1,-1)$ | $(1,0)$ | $(0,1)$ |
|---|---|---|---|---|---|
| | | $\mathrm{HE}_{m+1,n}^{e}$ | $\mathrm{HE}_{m+1,n}^{o}$ | $\mathrm{OAM}_{+m,n}^{+1}$ | $\mathrm{OAM}_{-m,n}^{-1}$ |
| $(1,1)$ | $\mathrm{EH}_{m-1,n}^{e}\ \mathrm{TM}_{0n}$ $(m\geq 2)\ (m=1)$ | $\mathrm{LP}_{mn}^{e,x}$ | $\mathrm{HP}_{mn}^{e,+1}$ | | |
| $(1,-1)$ | $\mathrm{EH}_{m-1,n}^{o}\ \mathrm{TE}_{0n}$ $(m\geq 2)\ (m=1)$ | $\mathrm{HP}_{mn}^{e,-1}$ | $\mathrm{LP}_{mn}^{o,x}$ | | |
| $(1,0)$ | $\mathrm{OAM}_{+m,n}^{-1}$ | | | $\mathrm{OAM}_{+m,n}^{x}$ | $\mathrm{CL}_{m,n}^{e,-1}$ |
| $(0,1)$ | $\mathrm{OAM}_{-m,n}^{+1}$ | | | $\mathrm{CL}_{m,n}^{e,+1}$ | $\mathrm{OAM}_{-m,n}^{x}$ |

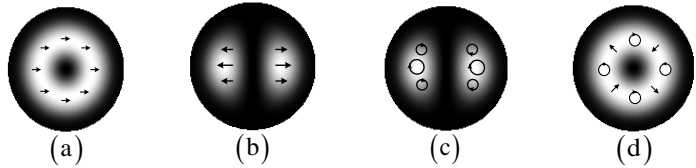

(a)　(b)　(c)　(d)

Fig. 1. Mode field distributions with polarization states, (a) $\mathrm{OAM}_{\pm 1,1}^{x}$, (b) $\mathrm{LP}_{11}^{e,x}$, (c) $\mathrm{CL}_{11}^{e,+1}$, (d) $\mathrm{HP}_{11}^{e,-1}$.

To intuitively visualize these combined modes based on vector modes with mode order being $m=n=1$, we give the mode field distributions of linear polarized OAM mode, conventional LP mode, CL and HP modes in Fig. 1. As for the odd, even and $y$-polarization counterparts that are not listed in Tab. 1. Actually, one can obtain them by controlling the initial phase of OAM mode bases.

## 3. Mode conversion using superposed helical gratings

HGs have been studied on OAM conversion in high-contrast-index RCFs where independent OAM modes can transmit without degeneration [20, 24, 25]. However, in the conventional FMFs belonging to WGFs, the circularly polarized OAM modes are vulnerable to degeneration into the linearly polarized OAM, CL, LP or HP modes, because of nearly the same effective index between two OAM modes with spin-orbit aligned and dis-aligned states, respectively. In this section, we theoretically study the mode coupling and combination principle of SHGs with opposite orientation and use them to generate vector modes in RCFs where the vector modes can independently supported without mode degeneracy [16, 19]. Furthermore, we also give the investigation of generation of linearly polarized OAM modes, CL and LP modes in the conventional FMFs, which belongs to degenerated modes.

In our discussion and presentation below, we supervise the OAM and SAM states when mode interacts with SHGs. Before that, we need to further simplify the expression forms of OAM modes and SHGs, as well as the coupling rules of HGs. The fiber-guided OAM modes as mode bases propagating along the $z$ direction can be written as,

$$\text{OAM}_{m,n}^{s} = F_m(r) e^{i(m\phi - \beta_{s,m} z)} \begin{bmatrix} 1 \\ s \cdot i \end{bmatrix}, \qquad (22)$$

where $m$ indicates the topological charge of OAM mode, and $s$ is associated with SAM, $s = -1$, $+1$ or $0$ correspond to right circular, left circular or linear polarization, respectively, $\omega$ is angular frequency, $\beta_{s,m}$ represents the propagation constant of the OAM mode, defined by $\beta_{s,m} = 2\pi n_{eff}/\lambda$ with $n_{eff}$ being the effective refractive index of corresponding vector mode in fibers. For the sake of simplicity, we use the denotation $|s,m\rangle$ to describe the angular momentum states of OAM modes.

The modulation function of a $l$-fold HG inscribed in fiber core can be roughly described as

$$\delta n(r,\phi,z) = \Delta n \rho(r) \cos\left[l\left(\phi + \sigma \frac{2\pi}{\Lambda} z\right)\right], \quad r \in \text{core region}, \qquad (23)$$

where $\Delta n$ is the modulation strength, $l$ is positive integers indicating the fold number of helical fringes of HGs, $\Lambda$ denotes grating period, $\sigma$ the helix orientation or handedness, $\sigma = +1$ and $-1$ correspond to left-handed and right-handed helix, respectively. $\rho(r)$ determinates the saturability of index modulation in the core region defined by $\rho(r) = r^2/a^2$ with $a$ being the radius of fiber core. Analogous to the description of OAM modes, the HGs here can be characterized by the denotation $[\sigma l, \Lambda]$. As for this kind of HGs, it may be fabricated by rotating the fiber when writing them by point-by-point method or interference beam with ultraviolet light in about one micrometer range [25-27].

Following the coupling rules of HGs in the reference [20], we unify the expression forms of OAM interaction with HGs as follows

$$|s,j\rangle \xrightarrow{[\sigma l, \Lambda]} |s,k\rangle \quad (j \neq k), \qquad (24)$$

for the transmission HGs, where

$$\sigma = \text{sign}\left[\left(\beta_{s,k} - \beta_{s,j}\right) \cdot (j-k)\right], \qquad (25)$$

$$l = |k - j|. \qquad (26)$$

$$\Lambda = 2\pi \left| \frac{j-k}{\beta_{s,k} - \beta_{s,j}} \right|, \tag{27}$$

and

$$|s, j\rangle \xleftrightarrow{[\sigma l, \Lambda]} |-s, k\rangle \quad (j \neq -k), \tag{28}$$

for reflection HGs, where

$$\sigma = \text{sign}\left[-\left(\beta_{-s,k} + \beta_{s,j}\right) \cdot (j+k)\right], \tag{29}$$

$$l = |j+k|, \tag{30}$$

$$\Lambda = 2\pi \left| \frac{j+k}{\beta_{-s,k} + \beta_{s,j}} \right|. \tag{31}$$

The coupling rules of OAM mode conversion above are based on the classic phase matching condition of fiber gratings and the specific helix matching condition being the requirement for OAM coupling of HGs.

When superposing the HGs with the identical gratings periods at the same region of fiber core, we can describe the index modulation as

$$\delta n(r,\phi,z) = \Delta n \rho(r) \left\{ \cos\left[l\left(\phi + \sigma \frac{2\pi}{\Lambda} z\right)\right] + \cos\left[l\left(\phi - \sigma \frac{2\pi}{\Lambda} z\right) + \Delta\varphi\right] \right\}, \quad r \in \text{core region}, \tag{32}$$

where $\Delta\varphi$ denotes the phase difference between two HGs components with opposite helix orientations. The schematic diagrams of SHGs $[-1,\Lambda] + \exp(i\Delta\varphi) \cdot [+1,\Lambda]$ and RCF are shown in Fig. 2. Fig. 2(a) and Fig. 2(b) correspond to the case of $\Delta\varphi = 0$ and $\Delta\varphi = \pi/2$, respectively. The cross-section and refractive index distribution of of RCF is shown in Fig. 2(c). Each helical modulation component of SHGs can independently modulate OAM of propagating modes into other OAM states. The linear superposition of modulated OAM states in the output of SHGs could produce a new modal forms. This is our main goal to achievement of different modal forms conversion using SHGs in this paper.

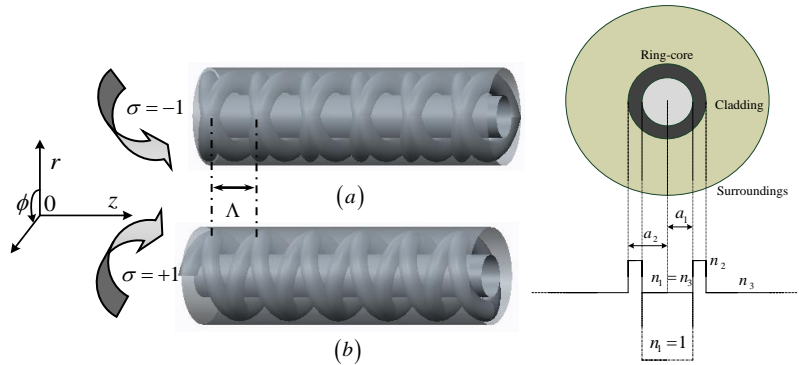

Fig. 2. The diagram of SHGs $[-1,\Lambda] + \exp(i\Delta\varphi) \cdot [+1,\Lambda]$ formed in the ring core of RCF, (a), $\Delta\varphi = 0$ and (b), $\Delta\varphi = \pi/2$. (c) Cross-section and refractive index distribution of of RCF.

Firstly, we give the coupling rules of pure OAM mode conversion, which can be independently generated from the circularly polarized fundamental mode according to Eqs. (24)-(31), as follows

$$\mathrm{HE}_{11}^{\pm 1}|\pm 1,0\rangle \xleftrightarrow{[\pm l,\Lambda_1]} \mathrm{OAM}_{\pm l,1}^{\pm 1}|\pm 1,\pm l\rangle, \tag{33}$$

$$\mathrm{HE}_{11}^{\pm 1}|\pm 1,0\rangle \xleftrightarrow{[\mp l,\Lambda_2]} \mathrm{OAM}_{\mp l,1}^{\pm 1}|\pm 1,\mp l\rangle \quad (l \neq 1), \tag{34}$$

where $\Lambda_1$ is the grating period associated with the effective index difference between the fundamental mode and spin-orbit aligned modes, whereas $\Lambda_2$ is that between the fundamental mode and spin-orbit dis-aligned modes, which is determined by Eq. (27).

If the incidence is linearly polarized Gauss beam, the excited linearly polarized fundamental mode $\mathrm{HE}_{11}$ will be converted into vector modes $\mathrm{HE}$, $\mathrm{EH}$ $(l>1)$, $\mathrm{TM}_{01}$ and $\mathrm{TE}_{01}$ by SHGs with a phase difference $\Delta\varphi$ that determines the azimuthal dependence of vector fields, i.e.,

$$\mathrm{HE}_{11}(|+1,0\rangle+|-1,0\rangle) \xleftrightarrow{[+l,\Lambda_1]+\exp(i\Delta\varphi)\cdot[-l,\Lambda_1]} \begin{cases} \mathrm{HE}_{l+1,1}^{e}(|-1,-l\rangle+|+1,+l\rangle), & \Delta\varphi = m\pi \\ \mathrm{HE}_{l+1,1}^{o}(|-1,-l\rangle+|+1,+l\rangle), & \Delta\varphi = \left(m+\frac{1}{2}\right)\pi \end{cases}, \tag{35}$$

$$\mathrm{HE}_{11}(|+1,0\rangle+|-1,0\rangle) \xleftrightarrow{[+l,\Lambda_2]+\exp(i\Delta\varphi)[-l,\Lambda_2]} \begin{cases} \mathrm{EH}_{l-1,1}^{e}(|-1,+l\rangle+|+1,-l\rangle), & \Delta\varphi = m\pi \\ \mathrm{EH}_{l-1,1}^{o}(|-1,+l\rangle+|+1,-l\rangle), & \Delta\varphi = \left(m+\frac{1}{2}\right)\pi \end{cases}, \tag{36}$$

$$\mathrm{HE}_{11}(|+1,0\rangle+|-1,0\rangle) \xleftrightarrow{[+1,\Lambda_3]+[-1,\Lambda_3]} \mathrm{TM}_{01}|0,0\rangle, \tag{37}$$

$$\mathrm{HE}_{11}(|+1,0\rangle+|-1,0\rangle) \xleftrightarrow{[+1,\Lambda_4]+[-1,\Lambda_4]} \mathrm{TE}_{01}|0,0\rangle, \tag{38}$$

where $m$ is non-negative integer, $\Lambda_3$ and $\Lambda_4$ are connected with the effective index difference between the fundamental mode and $\mathrm{TM}_{01}$ modes and $\mathrm{TE}_{01}$ modes, respectively. In Fig. 3. we show the sketch of mode conversion from the fundamental mode to vector modes $\mathrm{HE}_{21}^{e}$ by SHGs with $\Delta\varphi = 0$ in RCF, and particularly indicate the transformation of SAM and OAM states through SHGs. We use the blue helical phase with handedness to describe the OAM, and the red circles with direction indication to the SAM. The linear polarization can be divided into the components of right-hand and left-hand circular polarization that carries SAM of $-1$ and $+1$, respectively. The RCF parameters are designed as: inside and outside radii of ring-core waveguide are $a_1 = 3.00$ μm, $a_2 = 4.40$ μm, respectively, and the refractive index of innermost layer and equivalently infinite cladding are set as $n_1 = n_3 = 1.62$, that of ring-core waveguide is $n_2 = 1.64$. The grating parameters are designed as $\Lambda_1 = 807.4$ μm, $L = 3.0$ μm, and $\Delta n = 4.2 \times 10^{-5}$. The grating period is determined by the Eq. (27), and the grating length is by the coupling coefficient $\kappa$ associated with the full coupling condition of $\kappa L = \pi/2$ [20]. We numerically simulate the spectrum of power exchange for the mode coupling, and the results are shown in Fig. 4, where appear three resonance peaks corresponding to vector modes $\mathrm{TM}_{01}$, $\mathrm{HE}_{21}$ and $\mathrm{TE}_{01}$, respectively. The these peaks is split because of the effect of modal dispersion in this RCF for these vector modes. The conversion bandwidth can be adjusted by controlling the grating length during the fabricating process of these HGs to change the number of grating periods. The coupling efficiency from $\mathrm{HE}_{11}$ mode to $\mathrm{TM}_{01}$, $\mathrm{HE}_{21}$ and $\mathrm{TE}_{01}$ modes is different because of the not identical coupling coefficients between $\mathrm{HE}_{11}$ and these vector modes.

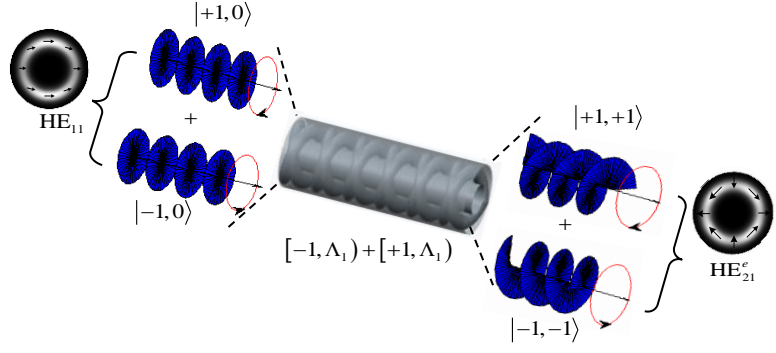

Fig. 3. The sketch of mode conversion from the fundamental modes $HE_{11}$ to vector modes $HE_{21}$ by SHGs $[-1,\Lambda_1)+[+1,\Lambda_1)$ with $\Lambda_1=807.4$ μm in RCF.

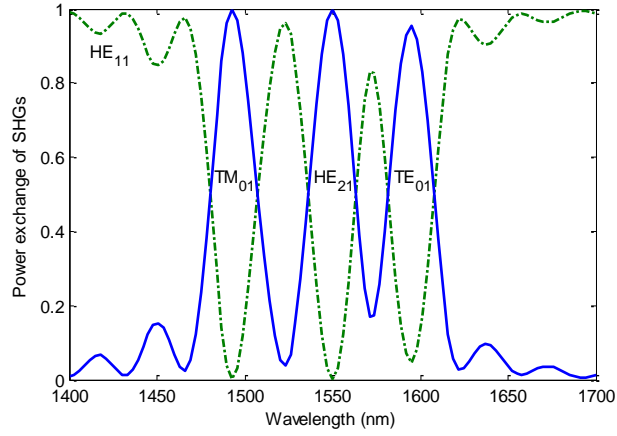

Fig. 4. The spectrum of power exchange for the coupling by SHGs $[-1,\Lambda_1)+[+1,\Lambda_1)$ with $\Lambda_1=807.4$ μm, $L=3.0$ cm, and $\Delta n=4.2\times10^{-5}$.

However, in the WGFs, such as the conventional FMFs, the vector modes would undergo degeneration into LP or HP modes due to the nearly same values of effective index, whereas the circularly polarized OAM modes would become linearly polarized OAM or CL modes, as shown in Tab. 1. As for this case, these three peaks in Fig. 4 would merge into one peak at one fixed wavelength, this is to say, the vector modes $TM_{01}$, $HE_{21}$ and $TE_{01}$ form the degenerated modes, such as LP, CL, and linearly polarized OAM modes due to the weakly guiding condition. The output modal forms are determined by the input polarization and the phase difference of SHGs.

In a FMF with parameters: the core and cladding radii $a_1=6.00$ μm, $a_2=62.50$ μm, respectively, and the refractive index of core and cladding $n_1=1.626$ and $n_2=1.62$, respectively. The normalized waveguide frequency is $V=3.4$ (at 1550nm), and thus support $LP_{01}$ and $LP_{11}$ modes, but $LP_{21}$ modes and higher order LP modes are cutoff. When the $LP_{01}$ mode just interacts with an uniform HGs [20], instead of SHGs, a linearly polarized OAM modes can be obtained, but the polarization states remain unchanged, as follows:

$$LP_{01}^{x,y}\begin{pmatrix}|+1,0\rangle\\+|-1,0\rangle\end{pmatrix}\xleftarrow{[\pm l,\Lambda]}OAM_{\pm l,1}^{x,y}\begin{pmatrix}|+1,\pm l\rangle\\+|-1,\pm l\rangle\end{pmatrix}. \quad (39)$$

If the incident fundamental mode has a circular polarization, through the SHGs with a phase difference $\Delta\varphi$, we will get the azimuth-dependent CL modes. As shown in Fig. 5, the fundamental mode with left-handed circular polarization is coupled into the azimuth-dependent $CL_{11}$ modes with the same polarization by the 1-fold SHGs. The grating parameters need to be designed as grating period $\Lambda = 625.0$ μm, and $\Delta\varphi = 0$ and $\pi/2$, correspondingly. The field distribution of even CL mode is characterized by a bilateral symmetry, and that of odd mode is by vertical symmetry, which coincides with the orientation of joint between two uniform HGs components with the opposite handedness.

$$HE_{11}^{\pm 1}(|\pm 1,0\rangle) \xleftrightarrow{[\pm l,\Lambda)+\exp(i\Delta\varphi)\cdot[\mp l,\Lambda)} \begin{cases} CL_{l1}^{e,\pm 1}(|\pm 1,\pm l\rangle+|\pm 1,\mp l\rangle) & \Delta\varphi = m\pi \\ CL_{l1}^{o,\pm 1}(|\pm 1,\pm l\rangle+|\pm 1,\mp l\rangle) & \Delta\varphi = \left(m+\frac{1}{2}\right)\pi \end{cases}. \qquad (40)$$

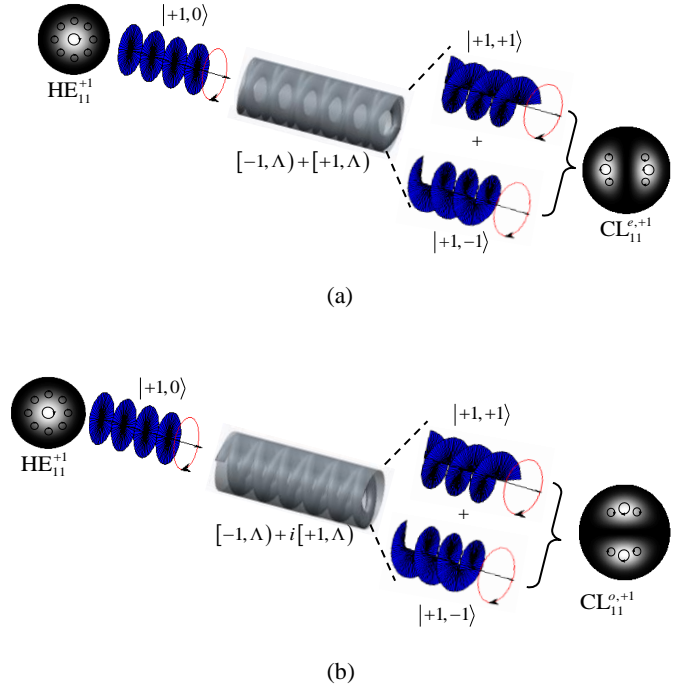

Fig. 5. The sketch of mode conversion in FMF from the left-handed circularly polarized fundamental modes $HE_{11}$ to modes $CL_{11}^{e,+1}$ by SHGs $[-1,\Lambda)+[+1,\Lambda)$ with $\Lambda = 625.0$ μm and $\Delta\varphi = 0$ (a), and modes $CL_{11}^{o,+1}$ by SHGs $[-1,\Lambda)+i[+1,\Lambda)$ with $\Lambda = 625.0$ μm and $\Delta\varphi = \pi/2$ (b).

Finally, under the incident condition of fundamental modes with linear polarization, we can generate LP modes by SHGs according to the coupling rules, as follows

$$LP_{01}^{x,y} \xleftrightarrow{[\pm l,\Lambda)+\exp(i\Delta\varphi)\cdot[\mp l,\Lambda)} \begin{cases} LP_{l1}^{e,x} & x, \Delta\varphi = m\pi \\ LP_{l1}^{o,x} & x, \Delta\varphi = \left(m+\frac{1}{2}\right)\pi \\ LP_{l1}^{e,y} & y, \Delta\varphi = m\pi \\ LP_{l1}^{o,y} & y, \Delta\varphi = \left(m+\frac{1}{2}\right)\pi \end{cases}. \qquad (41)$$

We show the conversion from the fundamental mode $LP_{01}$ with y-polarization to even $LP_{11}$ mode with the same polarization by SHGs with $\Delta\varphi = 0$ in Fig. 6. The $LP_{11}$ modes have four OAM mode components with different OAM and SAM states that stem from mode conversion from two circular polarization components of $LP_{01}$ mode. In theory, one can get desired higher order modes in different modal forms ($l > 1$), provided that the HGs or SHGs are modulated with more multiple fold numbers to satisfy the helix matching condition.

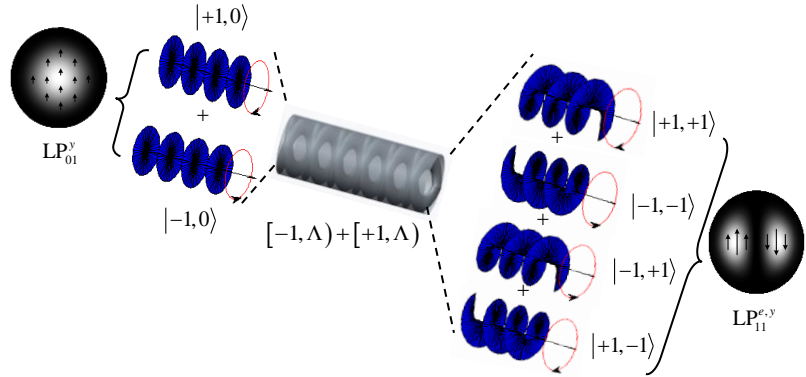

Fig. 6. The sketch of mode conversion in FMF from $LP_{01}^{y}$ to $LP_{11}^{e,y}$ by SHGs $[-1,\Lambda)+[+1,\Lambda)$ with $\Lambda = 625.0$ μm.

## 4. SUMMARY

In this article, we give detailed analysis and discussion on fiber-guided mode conversion between different modal forms. Various modal forms can be regarded as mode combination on the model basis of OAM modes. We reveal the correlation among them by the methods of Jones matrix. Based on the OAM mode conversion via uniform HGs, we propose the mode conversion from the fundamental mode to vector modes, LP modes, and CL modes utilizing SHGs. The investigation on mode correlation and conversion can make us well understood for fiber-guided modes and their relationship among different modal forms.

## ACKNOWLEDGMENT

I appreciate my supervisor Guoan Wu who gives me many opportunities and thank for every classmate in the institute of microwave technology and application. I also thank all the scholars who contribute to the improvement of the fiber gratings.

## References

1. X. Wang, Y. Li, J. Chen, C. Guo, J.Ding, and H. Wang, A new type of vector fields with hybrid states of polarization, Opt. Express **18**(2010)10786-10795.
2. A. Lizana, I. Estévez, F. A. Torres-Ruiz, A. Peinado, C. Ramirez, and J. Campos, Arbitrary state of polarization with customized degree of polarization generator, Opt. Lett. **40**(2015)3790-3793.
3. Q. Zhan, Cylindrical vector beams: from mathematical concepts to applications, Adv. Opt. Photonics **1**(2009)1-57.
4. K. Y Bliokh, A. Y. Bekshaev, and F. Nori, Dual electromagnetism: helicity, spin, momentum and angular momentum, New J. Phys. **15**(2013).033026.
5. L. Allen, M. W. Beijersbergen, R. J. C. Spreeuw, and J. P. Woerdman, Orbital angular momentum of light and the transformation of Laguerre-Gaussian laser modes, Phys. Rev. A **45**(1992)8185-8989.
6. K. Y. Bliokh, F. Nori, *Transverse and longitudinal angular momenta of light*, Phys. Rep., **592**(2015)1-38.
7. Y. I. Salamin, Electron acceleration from rest in vacuum by an axicon Gaussian laser beam, Phys. Rev. A **73**(2006)043402.

8. Q. Zhan, Trapping metallic Rayleigh particles with radial polarization, Opt. Express **12**(2004) 3377–3382.

9. M. Padgett, R. Bowman, Tweezers with a twist, Nat. Photonics **5**(2011)343-348.

10. M. Meier, V. Romano, and T. Feurer, Material processing with pulsed radially and azimuthally polarized laser radiation, Appl. Phys. A **86**(2007)329–334.

11. W. Chen and Q. Zhan, Three-dimensional focus shaping with cylindrical vector beams, Opt. Commun. **265**(2006)411–417.

12. J. Wang, J. Yang, I. M. Fazal, N. Ahmed, Y. Yan, H. Huang, Y. Ren, Y. Yue, S. Dolinar, M. Tur, and A. E. Willner, Terabit free-space data transmission employing orbital angular momentum multiplexing, Nature Photon. **6**(2012)488-496.

13. N. Bozinovic, Y. Yue, Y. X. Ren, M. Tur, P. Kristensen, H. Huang, A. E. Willner, and S. Ramachandran, Terabit-scale orbital angular momentum mode division multiplexing in fibers, Science **340** (2013)1545-1548.

14. J. Petersen, J. Volz, and A. Rauschenbeutel, Chiral nanophotonic waveguide interface based on spin-orbit interaction of light, Science **346**(2014)67-71.

15. J. Leach, B. Jack, J. Romero, A. Jha, A. M. Yao, S. Franke-Arnold, D. G. Ireland, R. W. Boyd, S. M. Barnett, M. J. Padgett, Quantum correlations in optical angle–orbital angular momentum variables, Science **329**(2010)662-664.

16. S. Ramachandran and P. Kristensen, Optical vortices in fiber, Nanophotonics **2**(2013)455-474.

17. A. W. Snyder and J. D. Love, *Optical Waveguide Theory,* Chapman and Hall, London, 1983.

18. K. Okamoto, *Fundamentals of Optical Waveguides*, (Academic, 2006), Chap. 3.

19. S. Ramachandran, P. Gregg, P. Kristensen. and S. E. Golowich, On the scalability of ring fiber designs for OAM multiplexing, Opt. Express **23**(2015)3721-3730.

20. L. Fang, and J. Wang, Flexible generation /conversion /exchange of fiber-guided orbital angular momentum modes using helical gratings, Opt. Lett. **17**(2015)4010-4014.

21. K. S. Lee, and T. Erdogan, Fiber mode conversion with tilted gratings in an optical fiber, J. Opt. Soc. Am. A **18**(2001)1176-1185.

22. L. Fang, H. Jia, H. Zhou, and B. Liu, Generation of cylindrically symmetric modes and orbital-angular-momentum modes with tilted optical gratings inscribed in high-numerical-aperture fibers, J. Opt. Soc. Am. A **32**(2015)150-155.

23. I. Moreno1, J. A. Davis, I. Ruiz, and D. M. Cottrell, Decomposition of radially and azimuthally polarized beams using a circular-polarization and vortex-sensing diffraction grating, Opt. Express **18**(2010)7173-7183.

24. L. Fang, and J. Wang, Mode Conversion and Orbital Angular Momentum Transfer Among Multiple Modes by Helical Gratings, IEEE J. Quantum Elect. **52** (2016) 6600306.

25. X. Zhang, A. Wang, R. Chen, Y. Zhou, H. Ming, and Q. Zhan, Generation and Conversion of Higher Order Optical Vortices in Optical Fiber With Helical Fiber Bragg Gratings, J. Lightwave Technol. 34(10) (2016)2413-2418.

26. R. J. Williams, R. G. Krämer, S. Nolte, and M. J. Withford, Femtosecond direct-writing of low-loss fiber Bragg gratings using a continuous core-scanning technique, Opt. Lett. 38 (11) (2013)1918-1920.

27. X. Zhang, A. Wang, R. Chen, Y. Zhou, H. Ming, and Q. Zhan , Generation and Conversion of Higher Order Optical Vortices in Optical Fiber With Helical Fiber Bragg Gratings, J. Lightwave Technol. 34(10) (2016)2413-2418.